\documentclass[aps,prc,twocolumn,superscriptaddress,showpacs,floatfix,nofootinbib]{revtex4-1}
\usepackage{amsmath,graphicx}
\def\bra{\langle}
\def\ket{\rangle}
\begin{document}

\title{Symmetric cumulants and event-plane correlations in Pb+Pb collisions}

\author{Giuliano Giacalone}
\affiliation{Institut de physique th\'eorique, Universit\'e Paris Saclay, CNRS,
CEA, F-91191 Gif-sur-Yvette, France} 
\author{Li Yan}
\affiliation{Institut de physique th\'eorique, Universit\'e Paris Saclay, CNRS,
CEA, F-91191 Gif-sur-Yvette, France} 
\author{Jacquelyn Noronha-Hostler}
\affiliation{Department of Physics, University of Houston, Houston TX 77204, USA}
\author{Jean-Yves Ollitrault}
\affiliation{Institut de physique th\'eorique, Universit\'e Paris Saclay, CNRS, CEA, F-91191 Gif-sur-Yvette, France} 
\date{\today}

\begin{abstract}
The ALICE Collaboration has recently measured the correlations between
amplitudes of anisotropic flow in different Fourier harmonics, referred to as symmetric cumulants. 
We derive approximate relations between symmetric cumulants involving $v_4$ and $v_5$ 
and the event-plane correlations measured by ATLAS. 
The validity of these relations is tested using event-by-event hydrodynamic calculations. 
The corresponding results are in better agreement with ALICE data than existing hydrodynamic predictions. 
We make quantitative predictions for three symmetric cumulants which are not yet measured. 
\end{abstract}

\maketitle

%\section{Introduction}
Anisotropic flow is the key observable showing that the matter produced in an ultrarelativistic nucleus-nucleus 
collision behaves collectively as a fluid~\cite{Heinz:2013th}. 
Following the discovery of flow fluctuations~\cite{Alver:2006wh} and 
triangular flow~\cite{Alver:2010gr}, a ``flow paradigm'' has emerged, which states that 
particles are emitted independently (up to short-range correlations) 
but with a momentum distribution that fluctuates event to event~\cite{Luzum:2011mm}. 
The azimuthal ($\varphi$) distribution in a given event is written as a Fourier series: 
\begin{equation}
\label{defVn}
P(\varphi)=\frac{1}{2\pi}\sum_{n=-\infty}^{+\infty}V_n e^{-in\varphi},
\end{equation}
where $V_n=v_n\exp(in\Psi_n)$ 
is the (complex) anisotropic flow coefficient in the $n$th
harmonic, and $V_{-n}=V_n^*$. 
Both the magnitude~\cite{Miller:2003kd} and phase~\cite{Andrade:2006yh,Alver:2006wh} of $V_n$ 
fluctuate event to event. 
In the last five years or so, an extremely rich phenomenology has emerged from this simple paradigm. 
RMS values of $v_n$
have been measured up to $n=6$~\cite{Adare:2011tg,ALICE:2011ab,ATLAS:2012at,Chatrchyan:2013kba}, 
and more recently, the full probability distribution of $v_n$~\cite{Aad:2013xma}. 
An even wider variety of new observables can be constructed by 
combining different Fourier harmonics~\cite{Teaney:2010vd,Bhalerao:2011yg,Bhalerao:2014xra}. 
This new direction was pioneered by the ALICE collaboration which
measured the angular correlation between $V_2$ and
$V_3$~\cite{ALICE:2011ab,Bhalerao:2011ry},  
and then explored systematically by the ATLAS collaboration which
analyzed fourteen mixed correlations
involving relative phases between Fourier harmonics, dubbed event-plane correlations~\cite{Aad:2014fla}. 

Recently, the ALICE collaboration has taken a new step in this direction~\cite{ALICE:2016kpq} 
by measuring the correlation between the magnitudes of different Fourier harmonics using a 
cumulant analysis~\cite{Bilandzic:2013kga}. 
We define the normalized symmetric cumulant 
$sc(n,m)$~\footnote{Note the ALICE collaboration uses the same notation for the numerator only.} with $n\not= m$ by
\begin{equation}
\label{defsc}
sc(n,m)\equiv\frac{\langle v_n^2v_m^2\rangle-\langle
  v_n^2\rangle\langle v_m^2\rangle}{\langle
  v_n^2\rangle\langle v_m^2\rangle}.
\end{equation}
ALICE has measured $sc(3,2)$ and $sc(4,2)$ as a function of centrality. 
While these two quantities are formally similar, the hydrodynamic mechanisms giving rise to these correlations differ. 
Elliptic flow, $v_2$, and triangular flow, $v_3$, are both determined to a good approximation by linear response 
to the anisotropies of the initial density profile in the corresponding harmonics~\cite{Niemi:2012aj,Gardim:2014tya}. 
Therefore, $sc(3,2)$ directly reflects correlations present in the initial spatial density profile, which are preserved by the 
hydrodynamic evolution as the spatial anisotropy is converted into a momentum anisotropy. 
Standard models for the initial density indeed reproduce the negative sign and overall (small) magnitude of the measured $sc(3,2)$ for all 
centralities~\cite{ALICE:2016kpq}. 
By contrast, $V_4$ gets a significant nonlinear contribution proportional to $V_2^2$ generated by the hydrodynamic 
evolution~\cite{Borghini:2005kd,Yan:2015jma,Qian:2016fpi} in addition
to the linear contribution from the initial anisotropy in the fourth  
harmonic~\cite{Gardim:2011xv,Teaney:2012ke}. 
The nonlinear response explains~\cite{Teaney:2013dta} the large event-plane correlation between $V_2$ and $V_4$. 
It also explains qualitatively why $sc(4,2)$ is positive. 

In this paper, we derive a proportionality relation between $sc(4,2)$
and the corresponding event-plane correlation, 
where the proportionality constant involves the fluctuations of
$v_2$. Using this, we are able to relate recent ALICE measurements with
previously measured quantities, which circumvents the most typical limitation of hydrodynamic 
predictions that depend on initial conditions or medium
properties~\cite{Luzum:2008cw,Gale:2012rq,Qiu:2012uy,Heinz:2013bua,Bozek:2013uha,vanderSchee:2013pia,Retinskaya:2013gca}.  
The sole assumption underlying our derivation is that the linear and
nonlinear contributions to $V_4$ are independent.  
The validity of this assumption is tested using hydrodynamic
calculations. 
The value of $sc(4,2)$ derived using our relation and previous ATLAS
measurements is compared with the recent direct measurement by ALICE.  
We make predictions along the same lines for 
$sc(5,2)$, $sc(5,3)$ and $sc(4,3)$, which are not yet measured.

%\section{Relating symmetric cumulants with event-plane correlations}
%\label{s:relations}

\begin{figure}[h]
\begin{center}
\includegraphics[width=0.55\linewidth]{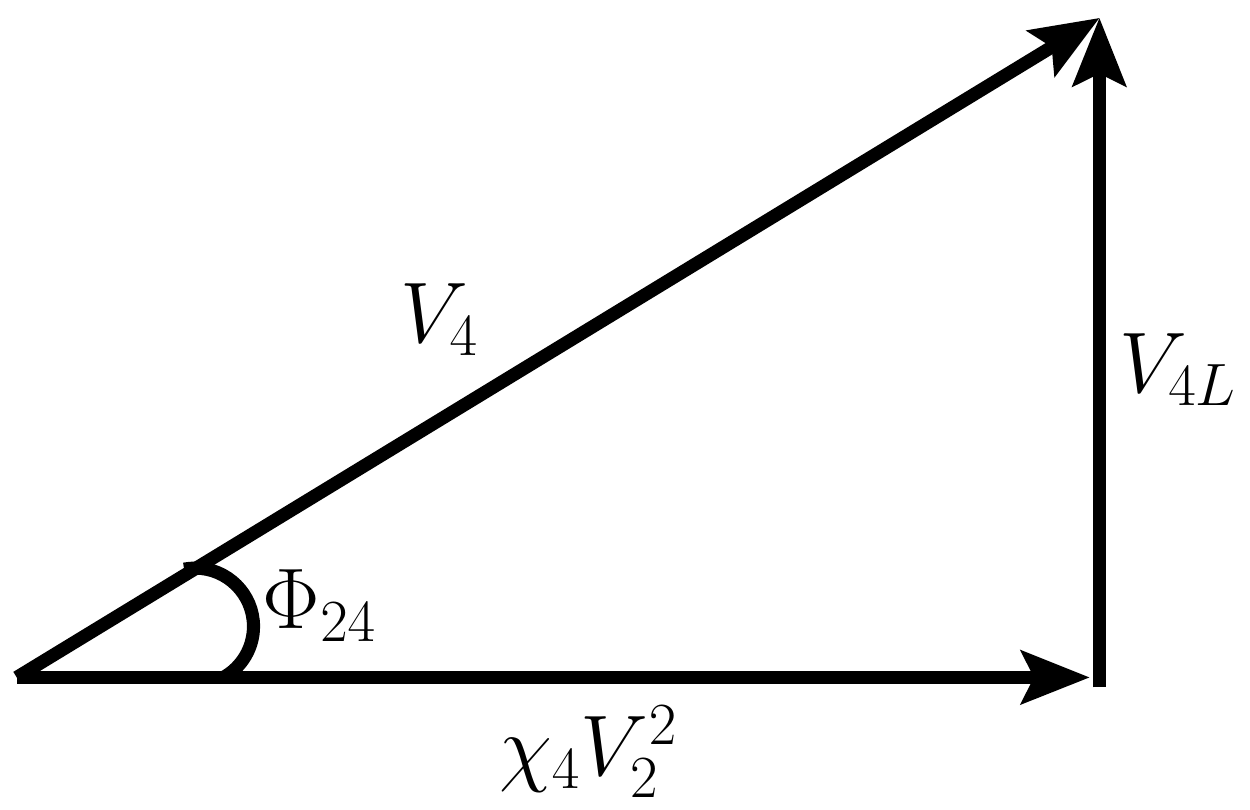} 
\end{center}
\caption{(Color online) 
\label{fig:triangle}
Schematic picture of the relation between the event-plane angle
$\Phi_{24}$  in Eq.~(\ref{pearson}) and the decomposition
Eq.~(\ref{decomposition}). The legs of the triangle correspond to the
rms values of the linear and nonlinear parts, and the hypothenuse is
the rms $v_4$.  A similar figure can be drawn for $V_5$. 
}
\end{figure}

We decompose $V_4$ and $V_5$ into linear and non-linear parts~\cite{Yan:2015jma}
\begin{eqnarray}
\label{decomposition}
V_4&=&V_{4L}+\chi_4 (V_2)^2\cr
V_5&=&V_{5L}+\chi_5 V_2V_3.
\end{eqnarray} 
We define $\chi_4$ and $\chi_5$ in such a way that the linear correlations 
between linear and nonlinear parts vanish, that is, 
$\langle V_{4L}(V_2)^{*2}\rangle=\langle V_{5L}V_2^{*}V_3^{*}\rangle=0$. 
We now introduce a measure of the relative magnitude of the linear and nonlinear parts via 
the Pearson correlation coefficients between $V_4$, or $V_5$, and their nonlinear parts:
\begin{eqnarray}
\label{pearson}
\cos\Phi_{24}
&\equiv& \frac{{\rm Re}\bra V_4 (V_2^{*})^2\ket}{\sqrt{\bra v_4^2
  \ket\bra v_2^4\ket}}\cr
\cos\Phi_{235}
&\equiv& \frac{{\rm Re}\bra V_5 V_2^{*}V_3^{*}\ket}{\sqrt{\bra v_5^2
  \ket\bra v_2^2v_3^2\ket}},
\end{eqnarray}
where $\Phi_{24}$ and $\Phi_{235}$ lie between $0$ and $\pi$. 
The first angle $\Phi_{24}$ corresponds precisely to the event-plane
correlation measured  
by ATLAS~\cite{Aad:2014fla} and denoted by 
$\langle\cos(4(\Phi_2-\Phi_4))\rangle_w$.\footnote{We only consider the event-plane correlations
  measured using the  scalar-product method, which are denoted by the
  subscript ``w'' in the ATLAS paper and have a  clear interpretation
  in terms of $V_n$, in contrast to the results obtained using the
  event-plane method~\cite{Luzum:2012da}.}
The second angle $\Phi_{235}$ almost corresponds to the quantity
denoted by $\langle\cos(2\Phi_2+3\Phi_3-5\Phi_5)\rangle_w$. 
The only difference is that the latter has 
$\langle v_2^2\rangle\langle v_3^2\rangle$ in the denominator, instead
of $\langle v_2^2v_3^2\rangle$~\cite{Yan:2015jma}. 
Therefore the precise relation is
\begin{equation}
\label{phi235}
\cos\Phi_{235}=\frac{\langle\cos(2\Phi_2+3\Phi_3-5\Phi_5)\rangle_w}{\sqrt{1+sc(3,2)}}, 
\end{equation}
where $sc(3,2)$ is defined in Eq.~(\ref{defsc}). 

Inserting Eq.~(\ref{decomposition})  into Eq.~(\ref{pearson}), one obtains
\begin{eqnarray}
\label{cos2}
\chi_4^2\bra v_2^4\ket &=&\bra v_4^2\ket \cos^2\Phi_{24}\cr 
\chi_5^2\bra v_2^2v_3^2\ket &=&\bra v_5^2\ket \cos^2\Phi_{235}.
\end{eqnarray}
These equations are exact and simply follow from the definition of
$\chi_4$ and $\chi_5$. They are depicted in Fig.~\ref{fig:triangle}.

\begin{figure}[h]
\begin{center}
\includegraphics[width=0.78\linewidth]{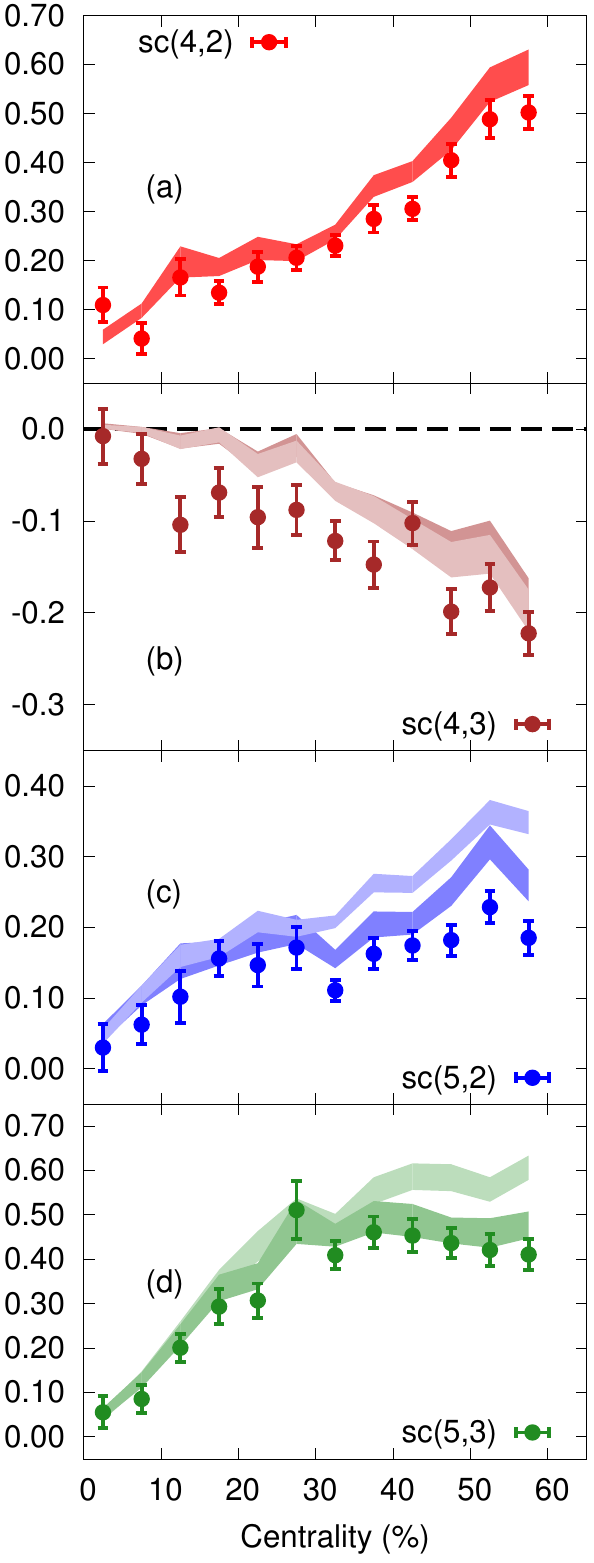} 
\end{center}
\caption{(Color online) 
\label{fig:hydro}
Test of Eqs.(\ref{exactrelations}) using hydro calculations. 
Symbols correspond to the left-hand sides of
Eqs.~(\ref{exactrelations}), dark shaded bands to the right-hand
sides. 
Light-shaded bands correspond to Eqs.~(\ref{approxrelations1}) and
(\ref{approxrelations2}). 
Errors are statistical and estimated via jackknife resampling. 
}
\end{figure} 
We now assume that the linear parts $V_{4L}$ and $V_{5L}$ are
statistically independent of $V_2$ and $V_3$. 
This is a stronger statement than just assuming that the linear
correlation vanishes. 
As will be shown below, it is a reasonable approximation in
hydrodynamics. 
Then, only the nonlinear response contributes to the correlation between $v_4$  and $v_2$, and  Eq.~(\ref{decomposition}) gives: 
\begin{equation}
\label{independence}
\langle v_4^2v_2^2\rangle-\langle v_4^2\rangle\langle
v_2^2\rangle=\chi_4^2 \left(\langle v_2^6\rangle-\langle
v_2^4\rangle\langle v_2^2\rangle\right).
\end{equation} 
Similar relations can be written for the correlations between $v_4^2$ and $v_3^2$,
$v_5^2$ and $v_2^2$ or $v_3^2$. 
Substituting in $\chi_4$ and $\chi_5$ extracted from
Eqs.~(\ref{cos2}), one obtains 
%\begin{eqnarray}
%\label{independence}
%\langle v_4^2v_2^2\rangle-\langle v_4^2\rangle\langle v_2^2\rangle&=&\chi_4^2 \left(\langle v_2^6\rangle-\langle v_2^4\rangle\langle v_2^2\rangle\right)\cr
%\langle v_4^2v_3^2\rangle-\langle v_4^2\rangle\langle v_3^2\rangle&=&\chi_4^2 \left(\langle v_2^4v_3^2\rangle-\langle v_2^4\rangle\langle v_3^2\rangle\right)\%cr
%\langle v_5^2v_2^2\rangle-\langle v_5^2\rangle\langle v_2^2\rangle&=&\chi_5^2 \%left(\langle v_2^4v_3^2\rangle-\langle v_2^2v_3^2\rangle\langle v_2^2\rangle\right)\cr
%\langle v_5^2v_3^2\rangle-\langle v_5^2\rangle\langle v_3^2\rangle&=&\chi_5^2 \left(\langle v_2^2v_3^4\rangle-\langle v_2^2v_3^2\rangle\langle v_3^2\rangle\right).
%\end{eqnarray} 
\begin{eqnarray}
\label{exactrelations}
sc(4,2)&=&\left(\frac{\bra v_2^6\ket}{\bra v_2^4\ket\bra v_2^2\ket}-1\right)\cos^2\Phi_{24}\cr
sc(4,3)&=&\left(\frac{\bra v_2^4v_3^2\ket}{\bra v_2^4\ket\bra v_3^2\ket}-1\right) \cos^2\Phi_{24}\cr
sc(5,2)&=&\left(\frac{\bra v_2^4v_3^2\ket}{\bra v_2^2v_3^2\ket\bra v_2^2\ket}-1\right)\cos^2\Phi_{235}\cr
sc(5,3)&=&\left(\frac{\bra v_2^2v_3^4\ket}{\bra v_2^2v_3^2\ket\bra v_3^2\ket}-1\right)\cos^2\Phi_{235}
\end{eqnarray}
These equations express symmetric cumulants in terms of event-plane
correlations and moments of $v_2$ and $v_3$.  
Based on these equations, one expects symmetric cumulants involving
$v_4$ or $v_5$ to increase with viscosity, in the same way as 
event-plane correlations~\cite{Teaney:2012gu,Niemi:2015qia}. 

In order to test Eqs.~(\ref{exactrelations}), we carry out event-by-event
hydrodynamic calculations using the same setup as in 
Ref.~\cite{Noronha-Hostler:2015dbi}:
initial conditions are given by the Monte-Carlo Glauber model~\cite{Alver:2008aq}, 
the shear viscosity over entropy ratio is $\eta/s=0.08$~\cite{Policastro:2001yc} within the viscous relativistic hydrodynamical model v-USPhydro \cite{Noronha-Hostler:2013gga,Noronha-Hostler:2014dqa}, 
and $V_n$ is calculated at freeze-out~\cite{Teaney:2003kp} for pions
in the transverse momentum range $0.2<p_t<3$~GeV/c. 
Note, however, that the particular setup used, and whether or not it
quantitatively reproduces experimental data, is irrelevant in this
context, since the statement is that Eqs.~(\ref{exactrelations})
should hold to a good approximation for {\it any\/} hydrodynamic
calculation. 
In hydrodynamics, $V_n$ can be computed exactly from the
one-particle momentum distribution for each
event~\cite{Schenke:2010rr,Gardim:2011qn,Qiu:2011iv}. 
%while only
%event-averaged quantities can be obtained experimentally due to the
%finite multiplicity. 
Therefore, reasonable accuracy is obtained with fewer events than in an
actual experiment. We generate 1000 events for each 5\% centrality bin. 
Figure~\ref{fig:hydro} displays the comparison between the left-hand
side (symbols) and the right-hand side (dark shaded bands) of
Eqs.~(\ref{exactrelations}). Agreement is good for all four quantities
and all centralities, in the sense that the absolute difference is
typically a few $10^{-2}$. 
The values of $sc(n,m)$ derived using Eqs.~(\ref{exactrelations}) are
in general above the actual values.\footnote{The difference between
  the two sides of  Eqs.~(\ref{exactrelations}) has the same sign for
  almost all centralities. The statistical error bar on this
  difference is in the range $0.02-0.03$, smaller than the difference
  itself.} 
This shows that the magnitude of of $V_{4L}$ (or $V_{5L}$) 
and that of $v_2$ (or $v_3$) are not quite independent in
hydrodynamics, but have a slight negative correlation. 
However, Eqs.~(\ref{exactrelations}) correctly capture the sign,
magnitude and centrality dependence of symmetric cumulants. 

\begin{figure}[h]
\begin{center}
\includegraphics[width=.9\linewidth]{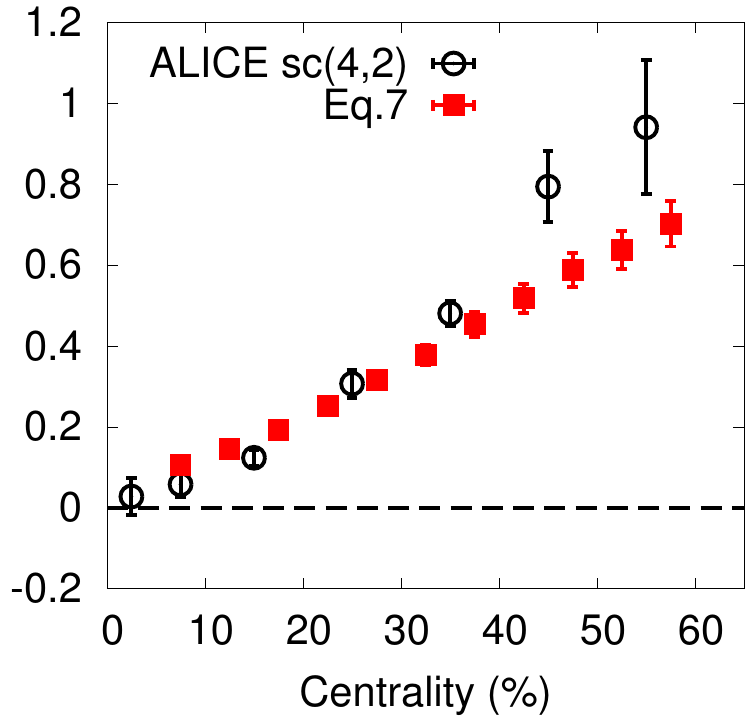} 
\end{center}
\caption{(Color online) 
\label{fig:sc42}
Open symbols: ALICE data for $sc(4,2)$~\cite{ALICE:2016kpq}. 
Closed symbols: value obtained using the right-hand side of 
Eq.~(\ref{exactrelations}) using ATLAS data for the moments of
$v_2$~\cite{Aad:2014vba} and the event-plane
correlation~\cite{Aad:2014fla}. 
}
\end{figure} 
The equation for $sc(4,2)$ can also be tested against existing data. 
The left-hand side has been measured by ALICE~\cite{ALICE:2016kpq} while 
the quantities entering the right-hand side (moments of the $v_2$ 
distribution, event-plane correlation) have been measured by
ATLAS. ALICE and ATLAS have different acceptances, both in
transverse momentum ($p_t$) and pseudorapidity ($\eta$), so that the
comparison is not quite apples to apples. However, we expect
that the quantities entering Eq.~(\ref{exactrelations}) (ratios of
moments, event-plane correlations) depend weakly on the $p_t$
range. The effect of the acceptance in $\eta$ will be discussed below. 
The moments of $v_2$ are not measured directly but 
can be expressed~\cite{Yan:2015jma} as a function of the cumulants
$v_2\{2\}$, $v_2\{4\}$ and $v_2\{6\}$, which are measured~\cite{Aad:2014vba}. 
Note that the ATLAS $v_2\{2\}$ is biased by nonflow correlations since
no rapidity gap is implemented ($v_2\{4\}$ and $v_2\{6\}$ are expected to
be free of nonflow correlations whether or not there is a gap). 
However, we have compared the ratios $v_2\{4\}/v_2\{2\}$ from
ALICE~\cite{ALICE:2011ab} (where $v_2\{2\}$ has a gap) and ATLAS
(where $v_2\{2\}$ has no gap) and found that they are compatible,
which suggests that the nonflow contribution to the integrated
$v_2\{2\}$ measured by ATLAS is small (nonflow effects are known to be
large at high $p_t$). 

Figure~\ref{fig:sc42} displays the comparison between the left-hand side
of Eq.~(\ref{exactrelations}) measured  by ALICE~\cite{ALICE:2016kpq}
and the right-hand side using ATLAS data. 
Agreement is reasonable for all centralities. In particular, 
our data-driven approach gives a better result for $sc(4,2)$
than existing hydrodynamic predictions~~\cite{ALICE:2016kpq,Niemi:2015qia}. 
%should we include results from Niemi in our plot?
Based on the hydrodynamic calculation of Fig.~\ref{fig:hydro}, one
would expect that the right-hand side of Eq.~(\ref{exactrelations}) is
larger than the left-hand side. However, it is the other way around 
above 30\% centrality. One reason may be that the event-plane
correlation for ATLAS uses a much larger pseudorapidity window
($|\eta|<4.8$) than ALICE ($|\eta|<0.8$). Now, the phase of $V_n$  
depends slightly on 
rapidity~\cite{Bozek:2010vz,Pang:2014pxa,Jia:2014ysa}, which induces a
decoherence of azimuthal correlations for larger $\Delta\eta$
\cite{Adamczyk:2013waa,Khachatryan:2015oea}. Due to these longitudinal
flow fluctuations, the event-plane correlation
measured by ATLAS is smaller than what ALICE would measure in a more
central rapidity window. Ideally, the comparison between the two sides
of Eq.~(\ref{exactrelations}) should be done in the exact same
rapidity window. 

We now make predictions for $sc(4,3)$, $sc(5,2)$ and $sc(5,3)$ using
Eqs.~(\ref{exactrelations}).  
The right-hand sides involve the mixed moments
$\langle v_2^4v_3^2\rangle$ 
and $\langle v_2^2v_3^4\rangle$ which could be measured
directly~\cite{Bhalerao:2014xra} but are not yet measured. 
However, the ALICE collaboration measures $|sc(3,2)|\ll 1$ for all
centralities~\cite{ALICE:2016kpq}, which implies $\langle
v_2^2v_3^2\rangle\approx \langle 
v_2^2\rangle\langle v_3^2\rangle$. Therefore, one can assume, as a
first approximation, that $v_2^2$ and $v_3^2$ are independent. 
For consistency's sake, we also neglect the correlation in evaluating 
$\Phi_{235}$, i.e., we make the approximation
$\cos\Phi_{235}\approx\langle\cos(2\Phi_2+3\Phi_3-5\Phi_5)\rangle_w$ 
(see Eq.~(\ref{phi235})). 
Eqs.~(\ref{exactrelations}) then give  
\begin{eqnarray}
\label{approxrelations1}
sc(5,2)&\approx&\left(\frac{\bra v_2^4\ket}{\bra v_2^2\ket^2}-1\right)
\langle\cos(2\Phi_2+3\Phi_3-5\Phi_5)\rangle_w^2\cr
sc(5,3)&\approx&\left(\frac{\bra v_3^4\ket}{\bra v_3^2\ket^2}-1\right)
\langle\cos(2\Phi_2+3\Phi_3-5\Phi_5)\rangle_w^2. 
\end{eqnarray}
The validity of Eqs.~(\ref{approxrelations1}) can again be tested using
event-by-event 
hydrodynamics. The right-hand sides are shown as light-shaded bands in
Figs.~\ref{fig:hydro} (c) and (d). Agreement is excellent for central
collisions but becomes worse as the centrality percentile increases,
as expected since we have neglected $sc(3,2)$ which becomes sizable
for peripheral collisions. 

\begin{figure}[h]
\begin{center}
\includegraphics[width=.9\linewidth]{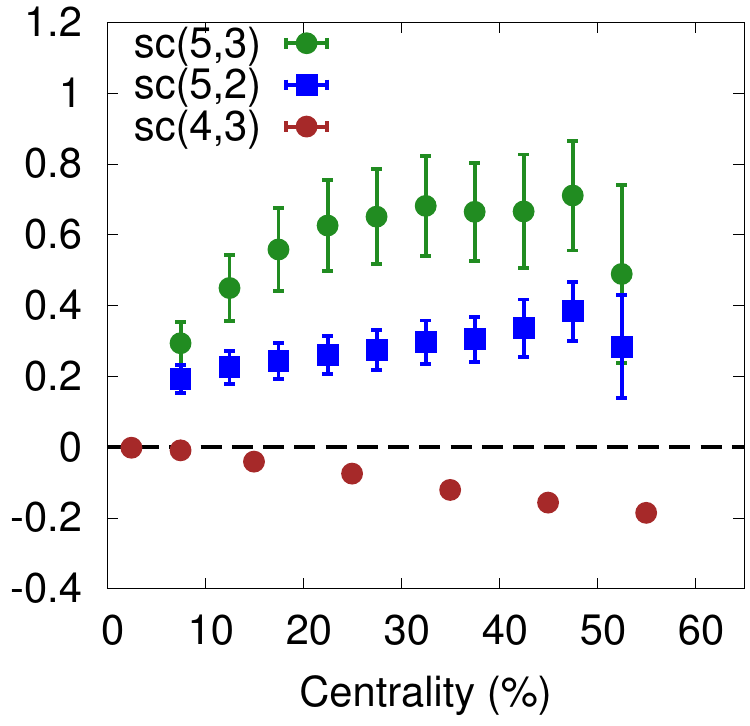} 
\end{center}
\caption{(Color online) 
\label{fig:prediction}
Predictions using the right-hand sides of 
Eqs.~(\ref{approxrelations1}) and (\ref{approxrelations2}), 
using ATLAS data for the moments of
$v_2$ and $v_3$~\cite{Aad:2014vba} and the event-plane 
correlations~\cite{Aad:2014fla}, and ALICE data for
$sc(3,2)$~\cite{ALICE:2016kpq}. 
}
\end{figure} 

If one assumes that $v_2^2$ and $v_3^2$ are independent, the second
line of Eqs.~(\ref{exactrelations}) gives $sc(4,3)=0$. 
In order to obtain a non-trivial prediction for
$sc(4,3)$, we need to take into account the 
small correlation between $v_2^2$ and $v_3^2$. 
We do this by assuming that $v_3^2$ can be decomposed as 
\begin{equation}
\label{v2v3}
v_3^2=cv_2^2+\beta,
\end{equation}
where $c$ is the same for all events in a centrality class,
and $\beta$ is independent of $v_2^2$.  
Using Eq.~(\ref{v2v3}), the correlation between an arbitrary moment of
$v_2$ and $v_3^2$ is given in terms of moments of $v_2$:
\begin{eqnarray}
\label{2eqs}
\langle v_2^2v_3^2\rangle-\langle v_2^2\rangle\langle v_3^2\rangle&=&
c\left(\langle v_2^4\rangle-\langle v_2^2\rangle^2\right)\cr
\langle v_2^4v_3^2\rangle-\langle v_2^4\rangle\langle v_3^2\rangle&=&
c\left(\langle v_2^6\rangle-\langle v_2^4\rangle\langle v_2^2\rangle\right).
\end{eqnarray}
The first equation relates $c$ with $sc(3,2)$ through
Eq.~(\ref{defsc}). 
Taking the ratio of Eqs.~(\ref{2eqs}) and inserting into 
Eq.~(\ref{exactrelations}), one obtains 
\begin{equation}
\label{approxrelations2}
sc(4,3)\approx\frac{\langle v_2^2\rangle\left(\langle v_2^6\rangle-\langle   v_2^4\rangle\langle   v_2^2\rangle\right)}
{\langle v_2^4\rangle\left(\langle v_2^4\rangle-\langle   v_2^2\rangle^2\right)}
sc(3,2)\cos^2\Phi_{24}.
\end{equation}
The right-hand side of this equation is shown as a light-shaded band
in Fig.~\ref{fig:hydro} (b). It is very close to 
the dark-shaded banded for all centralities, thus showing that the
decomposition in 
Eq.~(\ref{v2v3}) appropriately takes into account the correlation
between $v_2$ and $v_3$.  

Figure~\ref{fig:prediction} displays our predictions for $sc(5,3)$,
$sc(5,2)$ and $sc(4,3)$ using 
Eqs.~(\ref{approxrelations1}) and (\ref{approxrelations2}), 
where we use ATLAS data for the quantities in the right-hand side. 
For $sc(4,3)$, we use ALICE data for $sc(3,2)$, and the other
quantities in the right-hand side of Eq.~(\ref{approxrelations2}) 
(moments of $v_2$ and $\cos\Phi_{24}$) are
interpolated from ATLAS data, since ALICE and ATLAS use different
centrality bins. 

We have derived proportionality relations between  symmetric cumulants 
involving $v_4$ or $v_5$ and event-plane correlations. These relations
link correlations of different orders (symmetric
cumulants are 4-particle correlations, while event-plane correlations
are 3-particle correlations) and are fully non trivial. 
They are satisfied to a good approximation in event-by-event
hydrodynamics, and thus offer a direct test of hydrodynamic behavior,
which does not rely on a specific model of initial conditions 
and medium properties. 
The recent measurement of $sc(4,2)$ by ALICE passes the test. 
We have made predictions for $sc(5,2)$, $sc(5,3)$ and $sc(4,3)$ which
can be measured in the near future. These new observables will allow
to test hydrodynamic behavior directly, provided that one also
measures higher-order correlations between $v_2$ and $v_3$ such as 
$\langle v_2^4v_3^2\rangle$.

\section*{Acknowledgments}
This work is 
supported by the European Research Council under the
Advanced Investigator Grant ERC-AD-267258. JNH acknowledges the use of the Maxwell Cluster and the advanced support from the Center of Advanced Computing and Data Systems at the University of Houston to carry out the research presented here.

\end{document}